 \documentclass[12pt]{article}
\usepackage{amsfonts,amsmath}
\usepackage{amssymb}
\usepackage[hypertex] {hyperref}

\makeatletter \@addtoreset{equation}{section} \makeatother

\def\theequation{\thesection.\arabic{equation}}

\newcommand{\dis}{\displaystyle}

\newcommand{\ddd}{{\mathrm{d}}}

 \newcommand{\db}{ \mathbf{b}}
 \newcommand{\da}{\mathbf{a}}

\newcommand{\DDD}{{D}}

\newcommand{\dr}{{\rm d}}
\newcommand{\be}{ \begin{equation}}

\newcommand{\bp}{\bar{\partial}}

\newcommand{\NNN}{{\mathcal N}}

\newcommand{\III} {\mathcal{I}}%

\newcommand{\ee}{\end{equation}}
\newcommand{\PPP}{ \mathcal{{J}} }
\newcommand{\vf} {  {f}}

\newcommand{\bee}{\begin{eqnarray}}
\newcommand{\beee}{\begin{array}}
\newcommand{\eee}{\end{eqnarray}}
\newcommand{\eeee}{\end{array}}

\newcommand{\ga}{\alpha}

\newcommand{\pa}{{\ga^\prime}}

\newcommand{\pb}{{\gb^\prime}}
\newcommand{\pga}{{\gamma^\prime}}
\newcommand{\gb}{\beta}

\newcommand{\gga}{\gamma}

\newcommand{\M}{{\cal M}}

 \newcommand{\F}{{\cal F}}

\newcommand{\Hh}{{\cal H}}

\newcommand{\ie}{{\it i.e.,} }

\newcommand{\gd}{\delta}

\newcommand{\gvep}{\varepsilon}

\newcommand{\gs}{\sigma}

\newcommand{\go}{\omega}

\newcommand{\by}{{\bar{y}}}

\newcommand{\q}{\,,\qquad}
  \newcommand{\nn}{\nonumber}

\newcommand{\half}{\frac{1}{2}}
\newcommand{\ptl}{\partial}
\newcommand{\p}{\partial}
\newcommand{\D}{{\cal D}}
\newcommand{\f}{\frac}

 \newcommand{\bt}{\bar{t}}

\newsavebox{\ver}
\newsavebox{\verp}
\newsavebox{\gorp}
\newsavebox{\toch}

\begin{document}
\sbox{\ver}{\line(0,1){4}} \sbox{\gorp}{\line(1,0){7}}
\sbox{\verp}{\line(0,1){7}} \sbox{\toch}{\circle*{1}}

\begin{flushright}
\vspace{1mm}
 FIAN/TD/2015-11 \\
 October {2015}\\
\end{flushright}

\vskip1.5cm

 \begin{center}
 {\large\bf Symmetries of higher-spin current
 interactions in four dimensions}
 \vglue 0.6  true cm

\vskip0.5cm

 O.A. Gelfond$^1$ and M.A.~Vasiliev$^2$
 \vglue 0.3  true cm

 ${}^1$Institute of System Research of Russian Academy of Sciences,\\
 Nakhimovsky prospect 36-1,
 117218,
 Moscow, Russia

 \vglue 0.3  true cm

 ${}^2$I.E.Tamm Department of Theoretical Physics, Lebedev Physical
 Institute,\\
 Leninsky prospect 53, 119991, Moscow, Russia

 \end{center}

\vskip2cm

 \begin{abstract}
Current interaction of massless fields  in four dimensions is shown
to break  $ \mathfrak{sp} (8)$ symmetry of free massless equations
of all spins down to the conformal symmetry $ \mathfrak{su} (2,2)$.
This breaking is  in agreement with the form of  nonlinear
higher-spin field equations.

   \end{abstract}
\newpage


\section{Introduction}

The name of Igor Viktorovitch Tyutin is  well known worldwide
in the first place in relation with the BRST formalism.
Undoubtedly Tyutin is one of the leaders of Russian science, having made outstanding contribution to
relativistic quantum field theory far beyond the BRST formalism as such.
Working with Igor Viktorovitch in the same group for a long time we had
a great opportunity to fully appreciate both the true value of  his scientific potential
and the charm of his personality. For the volume  in honor of
 Igor   Tyutin's 75th birthday we are happy to contribute
  a paper where symmetries of relativistic systems are studied
  by the methods of  unfolded dynamics  having a much in common with
  the BRST approach  \cite{BRST}.

As shown by Fronsdal \cite{Fr1}, the infinite towers of free
massless fields that appear in the $4d$ higher-spin (HS) gauge
theory \cite{Gol} exhibit $\mathfrak{sp}(8)$ symmetry which  extends
the conformal symmetry $ \mathfrak{su}(2,2)$. The latter acts on
every  spin $s=0,1/2,1,3/2,2,\ldots $\,\,. The generators from
$\mathfrak{sp}(8)/s\mathfrak{u}(2,2)$ mix fields of different spins.

This observation suggests a manifestly $\mathfrak{sp}(8)$-symmetric
geometric realization of field equations of massless fields of
all spins  studied e.g. in   \cite{BL, BLS, BHS,Mar,Didenko:2003aa,
Plyushchay:2003tj,33,Fedoruk:2012ka,Bandos:2005mb,Florakis:2014aaa}
 (and references therein).
However attempts to extend this formalism to HS interactions were
not successful \cite{Bandos:2005mb}. The  full nonlinear
system of HS equations  \cite{more} possesses
 manifest $\mathfrak{sp}(4)$ symmetry rather than $\mathfrak{sp}(8)$.
It was not clear however whether this is an artifact of the formalism or
the $\mathfrak{sp}(8)$ symmetry is inevitably broken by interactions.

In this paper we show that HS current interactions necessarily
break $\mathfrak{sp}(8)$ down to its
conformal subalgebra  $\mathfrak{su}(2,2)$. Our analysis is based on
the approach of
\cite{GVC} where it was shown that current interactions can be
understood as a deformation of the two independent linear systems
for fields of rank-one and rank-two, associated, respectively, with
massless fields and conserved conformal currents in four space-time
dimensions. Though each of these systems is $\mathfrak{sp}(8)$ symmetric
it is not guaranteed that $\mathfrak{sp}(8)$ is preserved by the
deformation responsible for interactions.
Let us emphasise that our results do not rule out a possibility of introducing other
interactions, the existence of which is indicated by the results of  \cite{VZ,Florakis:2014aaa},
where   nontrivial
$\mathfrak{sp}(8)$ invariant three-point correlators were constructed.

The analysis of \cite{GVC} was  manifestly invariant under the $AdS_4$
symmetry algebra $ \mathfrak{sp} (4)$.  An important
question not considered in \cite{GVC} is to which extent the deformed
system respects $ \mathfrak{sp} (8)$. As  shown below,
in the interacting system,  $ \mathfrak{sp} (4)$ can be extended to the $4d$ conformal
algebra $ \mathfrak{su} (2,2)$ but not to the full  $ \mathfrak{sp}(8)$.
This implies that the current HS interactions break $ \mathfrak{sp}(8)$
down to its conformal subalgebra  $\mathfrak{su}(2,2)$.
Specifically, $ \mathfrak{sp} (8)$ contains the helicity operator
$\Hh$ such that its centralizer in $ \mathfrak{sp} (8)$ is
$\mathfrak{u}(2,2)$ spanned by the generators of conformal algebra $
\mathfrak{su} (2,2)$ and $\Hh$ itself. As we show, the $\mathfrak{u}(1)$
symmetry generated by $\Hh$
 cannot be extended to the mixed system.
Hence, the same is true for $\mathfrak{sp}(8)/\mathfrak{su}(2,2)$.

We analyze the system within the unfolded form of
\cite{GVC} in the sector of gauge invariant curvature 0-forms
which is free
of subtleties due to gauge symmetries. One of the main results of this paper
is that, rather unexpectedly,  this setup turns out to be manifestly conformal invariant.
The analysis in terms of 0-forms is more general than that in the
sector of gauge fields where the conformal invariance is   broken
for massless fields of spins $s>1$.  The latter breaking can be
resolved however on the $AdS$ background  \cite{33}.

The symmetry $\mathfrak{u}(1)\in \mathfrak{sp}(8)$ generated by the helicity operator
$\Hh$ describes electric-magnetic (EM)
 duality. Recall that the EM duality generalized to spins $s\ge1$ is generated by $\Hh$,
with respect to which (anti)self-dual solutions possess   eigenvalues $(-)s$.
In the case of $s=1$ this is the conventional EM duality.  Since current interactions are known to break
EM duality  our conclusions are not too surprising.
The breaking of $\mathfrak{sp}(8)$ considered in this paper occurs
in the sector of 0-forms and cannot be restored
via transition  to a curved background in  spirit of \cite{33}.

Our results are in agreement
with the form of nonlinear HS equations \cite{more} which also breaks
 EM duality. In fact, the form of nonlinear
HS equations suggests that this breaking is of spontaneous type
 {\it a la} embedding tensor formalism in supergravity
\cite{deWit:2002vt,de Wit:2007mt}.

In the rest of the paper we first recall relevant facts
of the unfolded dynamics formalism in Section 2. In Section
3 we reformulate the problem of current interactions
in terms of module deformation. The  conformally invariant
setup is worked out in Section 4. Symmetries of HS current interactions are
verified in Section 5. Obtained results are  briefly discussed in Section~ 6.

\section{Unfolded dynamics}
\label{Unfolded dynamics}
\subsection{Unfolded equations}
\label{Unfolded equationss}

Unfolded dynamics formalism
is most useful to control symmetries in a system.
The idea of this approach was suggested and applied to the
 interacting HS gauge theory in \cite{4dun,Ann}. For
more detail see also \cite{Gol,V_obz3,Didenko:2014dwa}.

Let $M^d$ be a $d$-dimensional manifold with coordinates $x^n$
($n = 0,1,\ldots d-1$).  Unfolded
formulation of a linear or nonlinear system of differential
equations and/or constraints  in $M^d$
assumes its  reformulation in the first-order form
\be
\label{unf} \dr  W^\Phi (x)= G^\Phi (W(x))\,,
\ee
where $\dis{ \ddd= d x^n
\frac{\p}{\p x^n}\,} $ is the exterior derivative on $M^d$,
$W^\Phi(x)$ is a set of degree-$p_\Phi$ differential forms and
$G^\Phi (W)$ is some degree $p_\Phi +1$ function of $W^\Phi$
\be\nn G^\Phi (W) = \sum_{n=1}^\infty
\vf^\Phi{}_{\Omega_1\ldots \Omega_n} W^{\Omega_1}\wedge \ldots \wedge
W^{\Omega_n}\,,
\ee
where $\vf^\Phi{}_{\Omega_1\ldots
\Omega_n}$  are  appropriately (anti)symmetrized structure
 coefficients and $G^\Phi$ satisfies the generalized Jacobi condition
\be \label{BI} G^\Omega (W)\wedge \f{\p G^\Phi (W)} {\p W^\Omega}
=0\,.
\ee

Strictly speaking, formal
consistency demands (\ref{BI})  be satisfied  at $p_{\Phi} <
d$ \,\, for a  $d$-dimen-sional manifold 
where any $d+1$-form is zero. Any solution of (\ref{BI})  defines a
 free differential algebra  \cite{FDA1,FDA2}.
The unfolded system is {\it universal} \cite{V_obz3} if the
generalized Jacobi identity is true independently of the dimension $d$,
\ie $W^\Phi$ can be treated as local coordinates of some target superspace.
Unfolded HS equations  including those considered in this paper are
universal.

For universal systems,  equation
(\ref{unf}) is invariant under the gauge transformation \be
\label{delw} \delta W^\Phi (x)= \dr  \varepsilon^\Phi (x) +
\varepsilon^\Omega (x)\wedge \frac{\p G^\Phi (W(x)) }{\p W^\Omega (x)}\,,
\ee where the gauge parameter $\varepsilon^\Phi (x) $ is a
  $(p_\Phi -1)$-form (which is zero for 0-forms $W^\Phi (x)$).

Local degrees of freedom are contained in the 0-form
sector. This is a consequence of the Poincar\'e lemma since
exact forms can be removed order by order by gauge transformation (\ref{delw}).
It should be noted that on the  language of unfolded equations
    local evolution of field
 is determined by values of $0$-forms at some fixed point of space-time . The transition to a
  standard
Cauchy problem is related to the decomposition of the fields into dynamical and auxiliary.
The latter
are expressed via derivatives of dynamical fields.  In addition, unfolded equations
impose differential equations on dynamical fields  like Klein-Gordon, Dirac
and other relativistic equations. Analysis of the dynamical content  of unfolded equations
is performed with the help of the    $\gs_-$-cohomology  technics of
 \cite{SVsc} (see also \cite{tens2}).%

Possible degrees of freedom of closed $p$-forms which can be
present due to nontrivial topology are not regarded as local.

A universal unfolded system can  be uplifted to a larger space via the
extension
\be
\ddd=dx^a\f{\p }{\p x^a} \quad\longrightarrow\quad
\hat{\ddd}=d\hat{x}^A \f{\p }{\p \hat{x}^A}= dx^a\f{\p }{\p x^a}+
dz^\ga\f{\p }{\p z^\ga}\,,
\ee
where $z^\ga$ are some additional
coordinates. An extension of a universal unfolded system in the
space $M$ with coordinates $x^a$ to a larger space $\M$ with
coordinates $\hat{x}^A$ remains formally consistent. Since the restriction
of the resulting system in $\M$ to the original system in $M$ is
achieved by restricting $\hat \ddd$ to $\ddd$, the local dynamical
contents of the two systems are equivalent. Indeed,
 initial data are still given by values of 0-forms at any point of
$\M$ which can be chosen to be in $M$. The  additional
equations in $\M$  reconstruct the dependence on $z$ in terms
of that in $x$. (Of course, this is true locally and the situation
can change if $z^\ga$ obey nontrivial boundary conditions.)
Allowing simple extensions to larger spaces,
universal unfolded systems provide  an efficient tool
 used in particular for the extension of the formulation of
massless fields from Minkowski space to the Lagrangian Grassmannian
in \cite{BHS} where the $\mathfrak{sp}(8)$ symmetry of HS multiplets acts geometrically.

\subsection{Vacuum}
\label{v}

The simplest class of universal  free differential algebras is in one-to-one correspondence
with Lie algebras. Indeed, let
 $w^\alpha$  be a set of  1-forms.
If no other forms are involved (e.g., all of them are consistently
set to zero in a larger system) the most general expression
for $  G^\alpha (\,w)$, that has to be a 2-form, is
$
G^\alpha (\,w)=-\half \vf^\alpha_{\beta\gamma}w^\beta\wedge w^\gga\,.
$
Consistency condition  (\ref{BI}) and unfolded
equations (\ref{unf}) impose, respectively, the
 Jacobi identity on the structure coefficients
$\vf^\alpha_{\beta\gamma}$ of a  Lie algebra $\mathfrak{g}$ and the flatness
condition on  $w^\ga$ \be \label{MC}
 \dr w^\alpha +\half
\vf^\alpha_{\beta\gamma}w^\beta\wedge w^\gamma=0\,. \ee

Transformation law  (\ref{delw}) yields the usual gauge
transformation of the connection $w$ \be \label{gw} \delta w^\ga (x)
= D\gvep^\ga (x):=
 \dr \gvep^\alpha (x)+\vf^\alpha_{\beta\gamma}w^\beta(x) \gvep^\gamma (x)\,.
\ee
A flat connection $w(x)$  is invariant under the global
transformations with the covariantly constant parameters \be
\label{glpar} D\gvep^\alpha (x)=0\,. \ee This equation is consistent
by virtue of
 (\ref{MC}). Therefore,  locally, it reconstructs
$\gvep^\alpha (x)$ in terms of its values $\gvep^\alpha (x_0)$ at
any given point $x_0$. $\gvep^\alpha (x_0)$ are the moduli of the
global symmetry $\mathfrak{g}$ that is now recognized as the
stability algebra of a given flat connection $w(x)$.

This example explains how
$\mathfrak{g}$-invariant vacuum fields appear in the unfolded
formulation.
Namely, vacuum is understood   as a solution of a nonlinear system possessing one or another
symmetry  $\mathfrak{g}$, which is usually   described by a
flat connection  valued in a representation of  $\mathfrak{g}$. %
 Typically, an unfolded system that contains 1-forms
$w^\ga$ associated with some Lie algebra $\mathfrak{g}$ admits a
flat connection $w^\ga$ as its  $\mathfrak{g}$-symmetric
vacuum solution.
In the perturbative analysis, the vacuum connection $w^\ga$ is assumed to
be a solution of some nonlinear system and to be of   order   zero.
 Such description of the background
geometry is coordinate independent. A manifest form of $w^\ga(x)$
is not needed unless one is interested in explicit solutions in the
specific coordinate system.

\subsection{Free fields and Chevalley-Eilenberg cohomology }
\label{ff}

Let us  linearize unfolded equation (\ref{unf}) around some
vacuum flat connection $w$  setting
 \be \label{lin1}
W^\Omega=w^\Omega+\omega^\Omega\,,
\ee
where $w$ is a flat connection of a Lie algebra $\mathfrak{g}$, which
solves (\ref{MC}), while  $\go^\Omega$ are
differential forms  of various degrees that are treated as small
perturbations and enter the equations linearly in the lowest order.
Consider first the
sector of forms $\go^i(x)$ of a given degree $p_i$ ({\it e.g.},
0-forms) within the set  $\go^\Omega(x)$. Then  $G^i$ is bilinear
in $w$ and $\go$, \ie $
 G^i =- w^\ga(T_\ga)^i {}_j \wedge \go^j.
$
 In this case   condition
(\ref{BI}) implies that the matrices $(T_\ga)^i {}_j$ form a
representation $T$ of $\mathfrak{g}$ in a vector space $V$ where
$\go^i(x)$  are valued. Corresponding equation
(\ref{unf}) is the covariant constancy condition \be \label{covc}
D_w \go^i=\dr\go^i+ w^\ga(T_\ga)^i {}_j \wedge \go^j=0 \ee
 with $D_w\equiv \dr+w$ being the covariant derivative
in the $\mathfrak{g}$-module $V$.

Equations (\ref{MC}) and (\ref{covc}) are invariant under
gauge transformations (\ref{delw}) with \be \label{delc} \delta
\go^i (x)= -\gvep^\ga(x)(T_\ga)^i {}_j  \go^j(x)\,. \ee Once the
vacuum connection is fixed,  system (\ref{covc}) is invariant
under the global symmetry $\mathfrak{g}$ with the parameters
satisfying (\ref{glpar}).

This simple analysis has useful consequences. First of all,
unfolding of any $\mathfrak{g}$-invariant linear  system of partial
differential equations implies its reformulated in terms of
$\mathfrak{g}$-modules.
Unfolded equations of the form (\ref{covc}) can be   integrated
in the pure gauge form
\be \label{pg} \go^i(x) = g^i{}_j (x,x_0)\go^j
(x_0)\q g^i{}_j (x,x_0)=P\exp-\int_{x_0}^x w^\ga T_\ga{}^i{}_j\,. \ee
If $\tilde{\mathfrak{g}}$  is a larger Lie algebra that acts in $V$,
$\mathfrak{g}\subset \tilde{\mathfrak{g}}\subset \mathfrak{gl}(V)$,
 it is also a symmetry of (\ref{covc}) simply because any
flat $\mathfrak{g}-$connection is the same time a flat
$\tilde{\mathfrak{g}}-$connection. As a result,
the Lie algebra $\mathfrak{gl}(V) $ of commutators of
endomorphisms of $V$ is the maximal symmetry of (\ref{covc}).

Let $\go^\da(x)$ and $\go^\db(x)$  be forms  of  different fixed
degrees $p_\da$ and $p_\db$, say, $p_\da -p_\db=k\geq0 $.
In the linearized approximation, one can consider
 $G^\Phi$  polylinear in the vacuum field
  $w^\ga$ but still linear in the dynamical fields $\go$
\be \label{co}
 G^\da (w, \go) =- \vf^\da_{\ga_1\ldots \ga_{k+1}\,,\db}
 w^{\ga_1}\wedge\ldots \wedge w^{\ga_{k+1}}\go^\db\,.
\ee

Let $\go^\db$ be a 0-form. The equation for $\go^\db$ is a
covariant constancy condition (\ref{covc}). The consistency
condition (\ref{BI}) applied to (\ref{co}) then literally implies
that $ \vf^\da_{\ga_1\ldots \ga_{k+1}\,,\db}w^{\ga_1}\wedge \ldots \wedge
w^{\ga_{k+1}} $ is a Chevalley-Eilenberg cocycle of $\mathfrak{g}$
with coefficients in $V_l \otimes V^*_r$ where $V_l$ is the module
where $G^\da $ is valued while $V^*_r$ is the module dual  to
that of $\go^\db$. Coboundaries are dynamically empty because, as is
easy to see, they can be removed by a field redefinition.
Thus, in the unfolded formulation, the Chevalley-Eilenberg
cohomology classifies possible nontrivial mixings between fields realized as
differential forms of different degrees \cite{33}. Specifically,
if both $\go^\da$ and $\go^\db$ are 0-forms, the  Chevalley-Eilenberg
cohomology (\ref{co}) describes a nontrivial deformation
of the direct sum of the modules associated with $\go^\da$ and
$\go^\db$. This case  is  most relevant to
the analysis of this paper where current interactions are described as a
 deformation of the direct sum of the modules associated with fields and
currents.

In presence of  deformation
(\ref{co}) the system remains invariant under the global symmetry
$\mathfrak{g}$ which is still the part of the gauge symmetry that
leaves invariant the vacuum fields $w^\ga$. Its action on $\go^\Phi$
is deformed however by (\ref{co}) according to (\ref{delw}).

\section{Higher-rank fields, nonlinear system and holography}

 Consider some $\mathfrak{g}$-symmetric free field-theoretical system. Its
unfolded formulation always contains a set of 0-forms (describing
matter fields and gauge
invariant combinations of derivatives of the gauge fields in the system) that form
a $\mathfrak{g}$--module $S$.
Being the space of all local degrees of freedom in the system, $S$ is closely
related to the Hilbert space $\mathrm{H}$ of single-particle states, that is a space
 of normalizable (positive-frequency) %
solutions of free equations  in the corresponding
free quantum theory  ($S$ is usually complex equivalent to $\mathrm{H}$ \cite{BHS}).
Then the analogue of the space of two-particle states is $Sym{S\otimes S} $ and
 of the space $M$ of all multiparticle states is
\be M(S)= \oplus_{r=0}^\infty S^r \q
S^r=Sym\underbrace{S\otimes\ldots \otimes S}_{r}.
  \ee
where $\DDD $ is a vacuum covariant derivative
in the $\mathfrak{g}$--module $S$.
Let 0-forms field   $C^i(x)$ of some unfolded system satisfy
 the vacuum covariant constancy condition
 \be \DDD  C^i(x)=0\,,
 \ee
where $\DDD $ is the vacuum covariant  derivative in $\mathfrak{g}$--module $S$.
Let $C^i(x)$ be called rank-one field.
Then a rank-$r$ field $C^{i_1\ldots i_r}(x)$ is defined to be
valued in  $S^r$ and satisfy the equation
\be \DDD_r{}  C^{i_1\ldots i_r}(x)=0\,,
\ee where $\DDD_r{} $ is
the vacuum covariant derivative of  $\mathfrak{g}$ in $S^r$. According to
Section \ref{ff}, the maximal symmetry of the rank-$r$
equation is $\mathfrak{gl}(S^r)$.

A basis of $S^r$ is formed by {\it pure} fields
being products of the rank-one fields
\be
\label{rp}  A_{i_1,\ldots i_r} C^{i_1}(x)\ldots
C^{i_r} (x)\,.
\ee
(Note that, unlikely to the bilocal
approach of \cite{Das:2003vw,Koch:2010cy}, the product is taken at
the same $x$ but for different values of the indices $i_k$ which
label a basis of $S$.) The space $S^r$ is invariant under
$\mathfrak{gl}(S)$ principally embedded into
$\mathfrak{gl}(S^r)$, \ie acting on every field in
(\ref{rp}).

The unfolded  equations for
the gauge invariant field strengths of massless fields of all spins in
$4d$ Minkowski space are \cite{Ann}
\be \label{minun}
\DDD  C(y,\by|x) :=
dx^{\ga \pb}\left(\f{\p}{\p x^{\ga\pb}}  +\f{\p^2}{\p
y^\ga\p \by{}^{\pb}}\right)C(y,\by|x)=0\,.
\ee
 Here $y^\ga$ and
$\by^\pb$ are auxiliary  mutually conjugated commuting
coordinates carrying  two-component spinor indices
$\ga,\gb = 1,2$; $\pa,\pb =  {1},  {2}$ while $x
^{\ga\pb}$ are Minkowski coordinates in two-component spinor
notations.

The rank-$r$  generalization of (\ref{minun}) is
\be
\label{minunK}
\DDD_r{}  C^{(r)}(y,\by|x) :=
dx^{\ga \pb}\left(\f{\p}{\p x^{\ga\pb}}  +
\frac{\ptl^2}{\ptl y^\ga_i\ptl {\by}^\pb_j}\eta_{ij}
\right) C^{(r)}(y,\by|x)=0\,\q \ee
where  $C^{(r)}(y,\by|x)=C(y_1,\ldots,y_r,\by_1,\ldots,\by_r|x)$ and $ \eta_{ij} $ is some nondegenerate  $r \times r$ matrix. 
According to  Section \ref{Unfolded dynamics},
Eq.~(\ref{minunK}) is $\mathfrak{sp}(8)$ symmetric simply because
$\mathfrak{sp}(8)$ is generated by the operators
\be
\left \{Y_j^A\,,
\f{\p}{\p Y_j^B}\right\}\q
\frac{\ptl^2}{\ptl Y^A_i\ptl Y^B_j}\eta_{ij}\q Y^A_i\, Y^B_j \eta^{ij}\, \qquad
\Big(Y^A_j=(y^\ga_j, \by^\pb_j)\Big)
\ee
which act on the functions $C^{(r)}(y ,\by)$.

As shown in \cite{tens2}, the unfolded form of the current conservation
condition is just the rank-two system while usual currents result from
bilinear substitution (\ref{rp}).
The construction of currents in the $C C$ form in 4d Minkowski space was discussed in detail in
\cite{GSV}. %
 The full current deformation of the HS field equations  involves gauge
1-forms $\go$. The {\it r.h.s.s} of their field equations  contain
conserved currents analogously to the  stress tensor
on the {\it r.h.s.s} of the Einstein equations. It is this sector
that contains most of usual current interactions. The rank-one 0-forms $C$
describe gauge invariant combinations of space-time derivatives of
$\go$. Current deformation of the equations on the gauge fields
$\go$ induces the deformation of the equations on $C$. Though this part of
the deformation is less familiar, it is in a certain sense simpler being
free of  complications  due to gauge symmetries.
Hence, in this paper we focus on the sector of 0-forms.

In the noninteracting case the modules $C$ and $J$ for fields
and currents
obey independent unfolded field equations
\be
\label{free}
\DDD C=0\q   \DDD _2 J=0\, \quad
\ee
with covariant derivatives
  $D$ (\ref{minun}) and $D_2$  (\ref{minunK}).
  Both of these systems are $\mathfrak{sp}(8)$ invariant hence being
consistent for arbitrary flat $\mathfrak{sp}(8)$ connection in $\DDD =\dr+ w$.
Schematically, the interacting system has the form %
\be
\label{int}
\DDD  C=F(w, J) \q \DDD_2{}  J=0\,,
\ee
where $F(w, J)$ describes interactions
once $J$ is realized in terms of bilinears of $C$.
In \cite{GVC},  current interactions of $4d$ massless
fields were described as a $\mathfrak{sp}(4)$-invariant linear system that mixes
rank-one massless fields $C$
with rank-two current fields $J$.  Group-theoretically, the
unfolded system of \cite{GVC} describes a deformation of the direct
sum of $\mathfrak{sp}(4)$--modules $S$ of massless fields and $S^2$
of conserved currents to an indecomposable $ \mathfrak{sp}
(4)$--module $S^{1,2}$ associated with the  exact sequence\footnote{
In \cite{33}, the fields $C^i$ were interpreted as coefficients in $C=C^i t_i$ in a
module with basis elements $t_i$. As usual for dual modules, for this
convention the arrows have to be reversed.}
\be 0\longrightarrow S^2 \longrightarrow
S^{1,2} \longrightarrow S \longrightarrow 0\,.
\ee
Here $S^2$ is a submodule of $S^{1,2}$ and $S = S^{1,2}/S^2$.

In \cite{GVC} it was checked  that the proper
current deformation of the free field equations describes a $\mathfrak{sp}(4)$-module
where the action of the algebra $\mathfrak{sp}(4)\subset \mathfrak{sp}(8)$ is
associated with the $AdS_4$ symmetry.
However, the existence of the $\mathfrak{sp}(4)$ invariant deformation of
(\ref{free}) to (\ref{int}) does not imply that the deformed action of $\mathfrak{sp}(4)$
can be extended to  $\mathfrak{sp}(8)$, \ie it is not guaranteed that Eq.~(\ref{int}) can be
consistently formulated for an arbitrary flat
$\mathfrak{sp}(8)$ connection $w$.
 It can happen that deformation (\ref{int}) only makes sense for
connections $\go$ valued in some subalgebra $\mathfrak{h}$ of the symmetry
algebra $\mathfrak{g}$ of system (\ref{free}), forming
 an $\mathfrak{h}$-module but not a $\mathfrak{g}$-module. In this
paper we will show that the deformation remains consistent for the
$\mathfrak{su}(2,2)$ extension of $\mathfrak{sp}(4)$ but not beyond, \ie
the interactions preserve conformal symmetry but not full $\mathfrak{sp}(8)$.

A few comments are now in order.

The proposed construction not
only properly  describes the field equations of massless fields in
presence of currents but also gives a useful tool for
reconstruction of Greens functions (at least in
the gauge invariant sector of 0-forms). Indeed, in the 0-form
sector that mixes massless 0-forms $C$ with current 0-forms
$J$, the deformed field equation has the covariant constancy form
(\ref{covc}) for some representation of the Minkowski or $AdS_4$
symmetry. Hence, in the 0-form
sector, the deformed equations can be solved in the pure gauge form
(\ref{pg}). The representation is indecomposable, having  triangular
form since $J$ contributes to the equation for $C$ but not otherwise.
The operator $g^i{}_j$
(\ref{pg}) also has a triangular form, containing the
off-diagonal term that maps $J$ to $C$ but not the other way around.
Hence, given $J$, $C$ is reconstructed in the form
\be C=G(J)+C_0\,, \ee
where $C_0$ is an arbitrary solution of the undeformed
field equations for $C$. Clearly, $G$ is  Greens function that reconstructs
solutions of massless  equations via currents on their {\it r.h.s.}
(More precisely, this is gauge invariant Greens function that
reconstructs gauge invariant field strengths of massless fields.) It
would be interesting to apply this method for a practical computation.

The current deformation of free field
equations  derived from the perturbative expansion of the
nonlinear HS equations of \cite{Ann,more}
(see however \cite{Boulanger:2015ova,Skvortsov:2015lja} on possible subtleties of such a derivation)
has  more general form than (\ref{int}) containing also the
HS gauge connections $\go(y,\bar y |x)$
\be
\label{int1}
\DDD  C+\go\star C -C\star \tilde \go=F(w, C)   \q \DDD_2{}  J=0\,,
\ee\be
\label{dg}
D^{ad} \go +\go \star \go = G(w,C)\,,
\ee
where bilinear combinations of $C$ should be identified with the current $J$,
 $\star$ is the Weyl star product acting on functions of $y$ and $\bar y$,
$D^{ad}  $ is the adjoint covariant derivative, and
$\tilde \go(y,\bar y|x) := \go(-y, \bar y |x)$ 
(for more detail see \cite{Ann,Gol}). A simple but important fact, which follows
 from the analysis of \cite{Ann}, is that the $\go$-dependent
terms on the {\it l.h.s.} and the terms  on the {\it r.h.s.} of Eq.~(\ref{int1}) contribute to different sectors of the equations. Namely, let $s$ be the spin of the field $C$ in the first term of
(\ref{int1}) while $s_1$ and $s_2$ be spins of the field constituents of $J\sim CC$.
Then for $s\ne0$ the $F$-terms are non-zero at
\be
\label{sss}
s\geq s_1 +s_2\,.
\ee
In other words, the gauge invariant currents $J$ built from the 0-forms $C$
have spin $s$ obeying (\ref{sss}).
The $\go$-dependent terms in (\ref{int1}) with $\go$ and $C$ carrying spins $s_1$ and
$s_2$  contribute to the equation for the spin-$s$ field $C$ with $s< s_1 +s_2$.
Note that this conclusion is in agreement with the results of \cite{Smirnov}
where the currents with spins beyond the region (\ref{sss}) were built in terms
of the gauge connections $\go$. In Appendix we show  that the
$\go$-dependent terms in (\ref{int1}) do not contribute to the consistency
check of Eq.~ (\ref{int1}) in the sector (\ref{sss}).

In this paper we focus on the sector of fields
obeying (\ref{sss}),  discarding the $\go$-dependent terms.
Note that, beyond the lowest-order  deformation, all kinds
of terms are anticipated to be mixed nontrivially, \ie the contribution of
the gauge connection $\go$  cannot be discarded in any sector at the full nonlinear level.

The $\go$-independent part of deformation (\ref{int}) admits an interesting
interpretation from the $AdS_4 /CFT_3$ duality perspective.
As shown in \cite{Vasiliev:2012vf}, the boundary limit of the $4d$ 0-forms
$C(y,\bar y|x) $ gives $3d$ conformal currents ${\mathcal J}$ which are $3d$
rank-two fields. As a result,
the holographic dual of the $4d$ current $J$ bilinear in $C$ is ${\mathcal J} {\mathcal J}$.
Hence the holographic version of equation (\ref{int}) has the form
\be
\label{3dj}
{\mathcal D} {\mathcal J} = F(w, J({\mathcal J}))\,,
\ee
where ${\mathcal D}$ is the $\mathfrak{sp}(4)$ covariant derivative at the boundary.
For $F=0$ this gives the unfolded current conservation condition
at the boundary \cite{Vasiliev:2012vf}. For $F\neq 0 $, the current conservation
condition is deformed. In \cite{Maldacena:2012sf} such a deformation was associated
with ``slightly broken HS symmetry".

It should be stressed however that the deformed HS equations still respect global
HS symmetries.
The latter are neither broken nor even deformed, \ie the algebra of HS
transformations remains unchanged. What is deformed is the transformation law.
Indeed, the nonlinear HS equations can be linearized with respect to an arbitrary
HS connection $w(Y|x)$ obeying the flatness condition $\dr w +w\star w=0$.
Since the full nonlinear HS equations are formally consistent, their
perturbative expansion around $w$  is also consistent at any
order. Application of  (\ref{delw}) to the vacuum field
$w$ gives the transformation law for global $HS$ symmetries in a chosen
background. ($w$ can be set equal to its $AdS_4$ value
upon differentiation in (\ref{delw}).) Preservation of the global
HS symmetries at the nonlinear level is not too surprising being
analogous to the fact that a perturbative expansion of any general coordinate
invariant theory
around Minkowski background
preserves Poincar\'e symmetry.

So far our consideration was a
consequence of the previous work of \cite{Ann,GVC}. The new result of this paper is
that the manifest symmetry of equation (\ref{3dj}) is larger than the boundary
conformal symmetry $\mathfrak{sp}(4)$. Surprisingly it gets
enhanced to the $4d$ conformal symmetry $\mathfrak{su}(2,2)$. Note however that this is a
symmetry of equation (\ref{3dj}) but not necessarily of the operator algebra
of boundary currents.

\section{Background $\mathfrak{u}(2,2)$ connection}
\label{sp8}
As sketched in Section \ref{Unfolded dynamics},
a system is $\mathfrak{g}$-symmetric if its unfolded equations are formally
consistent for any flat connection $w$ of $\mathfrak{g}$.

In  terms of two-component spinors,
the $\mathfrak{su}(2,2)\sim \mathfrak{o}(4,2)$ connections   are
$
h^{\ga\pa},\, \omega_{\ga}{}^{\gb},\,
\overline{\omega}_{\pa}{}^{\pb},\, b$
and $f_{\ga\pa}.$
Extending $\mathfrak{su}(2,2)$ to $\mathfrak{u}(2,2)$ by a central helicity
generator  with the gauge connection $\tilde{b}$, the $\mathfrak{u}(2,2)$
flatness conditions read
\bee
\label{h4}
R^{\ga\pb}&:=&
\dr h^{\ga\pb} -\omega_\gga{}^\ga\wedge h^{\gga\pb}-
\overline{\omega}_\pga{}^\pb\wedge h^{\ga\pga} =0\,,
\\
\nn
R_{\ga\pb}&:=&
\dr f_{\ga\pb} +\omega_\ga{}^\gga\wedge f_{\gga\pb}+
\overline{\omega}_\pb{}^\pga\wedge f_{\ga\pga} =0\,,
\\
\nn
R_\ga{}^\gb&:=&
\dr \omega_\ga{}^\gb +\omega_\ga{}^\gga\wedge \omega_\gga{}^\gb -
 f_{\ga\pga}\wedge h^{\pga\gb}=0\,,
\\
\nn \overline{R}_\pa{}^\pb &:=& \dr \overline{\omega}_\pa{}^\pb
+ \overline{\omega}_\pa{}^\pga\wedge \overline{\omega}_\pga{}^\pb -
f_{\gga\pa}\wedge h^{\gga\pb}=0\,.  \eee
Lorentz connection is described by the traceless parts
$\go^{L}{}_{\ga}{}^\gb$ and $\overline{\go}^{L}{}_{ \pa}{}^\pb$ of
$\go_\ga{}^\gb$ and $\overline{\go}_\pa{}^\pb$, respectively,
while their traces are associated with the gauge
fields $\dis{}$
\be \label{btb} b= \half \big
(\go_\ga{}^\ga + \overline{\go}_\pa{}^\pa\big )\q \tilde{b}= \half
\big (\go_\ga{}^\ga -\overline{\go}_\pa{}^\pa\big )\,.
\ee
Note that since $b$  and $\tilde{b}$ are  one-component
\be \label{btb1}
 b\wedge b=\tilde{b}\wedge\tilde{b}=0\,.
\ee
The $\mathfrak{u}(2,2) $ invariant rank-$r$ unfolded equations
are \be\label{Dr0}
\DDD_{r }{}_{\mathfrak{u}}^{tw} C^{(r)}(y , \by |x)=0\q
\ee
\be
\label{Dtwsp8r}
    D_{r }{}_{\mathfrak{u}}^{tw}:=
\dr    -
 \go^L{}^{\ga\gb}y_ j{}_\ga\p^j_\gb-
 \overline{\go}^L{}^{\pa\pb}\bar{y}_j{}{}_\pa \bp^j_\pb
 + f_{\ga\pa}y_i{}^\ga{}\by{}_ j {}^\pa \eta^{ij}
+
h^{\ga\pa}\p^i_\ga\bp{}^j_\pa
\eta_{ij} + b \D_r
+  \widetilde{b}\Hh_r
\,,\ee
 where
\be\label{hel1} \Hh_r=\half
\big ( y_ j{}{}^\ga \p^j{}_\ga
-\bar{y}_j{}{}^\pa \bp ^j{}_\pa\big )  \ee is a {\it   rank -$r$\,\,   helicity operator},
\be \label{dilatrankr}\D_r=r+ \half \big ( y_ j{}{}^\ga \p^j{}_\ga
+\bar{y}_j{}{}^\pa \bp ^j{}_\pa\big )
\ee is a {\it   rank -$r$    dilatation operator} and\be \nn \rule{0pt}{20pt}
\eta_{ij}\eta^{kj}=\gd^k_i\q\p^j_\gb=\frac{\p }{\p y_j^\gb}\q
  \bp^j_\pa=\frac{\p }{\p \by_j^\pa}\q
  i,j,k=1,\ldots r\,.\qquad  
  \ee
 The rank-one equation $ D_{1 }{}_{\mathfrak{u}}^{tw}C(y,\bar y|x)=0$ describes the
$\mathfrak{u}(2,2) $ invariant $4d$ massless field equations in terms of
the generalized Weyl tensors ${} C(y,\by|x)$.

For general rank, the conformal dimension $\Delta_r$ of a field component equals to
the eigenvalue of $\D_r$.
 For a spin-$s$ primary field of rank-$r$  the conformal dimension is
\be\label{conwe}
\Delta_r= r+s\,.\ee

The  conformal  covariant derivative $D_r{}_{\mathfrak{su}}^{tw}$ coincides with
$D_{r }{}_{\mathfrak{u}}^{tw}$ (\ref{Dtwsp8r}) at $\widetilde{b}=0$
  \bee\label{Dtwcr}
D_{r }{}_{\mathfrak{su}}^{tw}=D_{r }{}_{\mathfrak{u}}^{tw}\Big|_{\widetilde{b}=0} \,.
\eee
Eigenvalues of $\Hh_r$ (\ref{hel1}) describe  helicities of field components.
At $r=1$, $\Hh=\Hh_1$    is the usual helicity operator.
Conformal algebra $\mathfrak{su}(2,2)$ is the subalgebra of
$\mathfrak{sp}(8)$ spanned by elements  commuting with
 $\Hh \in \mathfrak{sp}(8)$ (\ref{hel1}). More
precisely, the centralizer of $\Hh$ in $\mathfrak{sp}(8)$ is
$\mathfrak{su}(2,2)\oplus \mathfrak{u}(1)$ where $\mathfrak{u}(1)$
is generated by $\Hh$. Conformal algebra $\mathfrak{su}(2,2)$ and $\Hh$
act on states of  definite helicities while $\mathfrak{sp}(8)$
mixes different helicities.

The $AdS_4$ geometry is described by
the Lorentz connections $\go^L{}^{\ga\gb}$,
$\overline{\go}^L{}^{\pa\pb}$ and vierbein $e^{\ga\pa}$  of
 $\mathfrak{sp}(4,{\mathbb R})\subset \mathfrak{su}(2,2)
 \subset\mathfrak{sp}(8,{\mathbb R})$
 via the substitution
\be
\label{adsc}
 h^{\ga\pa}=  \lambda e^{\ga\pa} \q
 f_{\ga\pa} =   \lambda  e_{\ga\pa}\q b=\tilde{b}=0\,,
\ee
which gives
\bee\label{Dtwads}
 D_r{}_{ads}^{tw}:=
 \dr    -
 \go^L{}^{\ga\gb}y_ j{}_\ga\p^j_\gb-
 \overline{\go}^L{}^{\pa\pb}\bar{y}_j{}{}_\pa \bp^j_\pb
 + \lambda e_{\ga\pa}y_i{}^\ga{}\by{}_ j {}^\pa \eta^{ij}
+
\lambda e^{\ga\pa}\p^i_\ga\bp{}^j_\pa
\eta_{ij}\,.
\eee
Two-component indices are raised and lowered   by the symplectic
forms $\gvep_{\ga\gb}$ and $\gvep_{\pa\pb}$
\be \label{Cind} A_\gb
=A^\ga \gvep_{\ga\gb}\q A_\pb =A^\pa \gvep_{\pa\pb}\q A^\ga =A_\gb
\gvep^{\ga\gb}\q A^\pa =A_\pb \gvep^{\pa\pb}.\qquad
 \ee

\section{Symmetries of current interactions}
\label{globsym}
\subsection{Structure of the current deformation}

As  shown in \cite{GVC}, in the unfolded dynamics approach, interactions
of $4d$  massless fields of all spins with currents
result from the linear problem describing a mixing of rank-one (massless particles)
and rank-two (current) systems.

 To make contact with \cite{GVC}, we set $\eta_{ij}=\gd_{ij}$, $\eta^{ij}=\gd^{ij}$
 and  introduce the new variables
\be\label{compl}
\sqrt{2}y^\pm{}^\ga= y ^\ga_1\pm i {y}_2^\ga\q
\sqrt{2}\by^\pm{}^\pa= \by ^\pa_1\pm i {\by}_2^\pa\q
\p_\pm{}_\ga{}=\frac{\partial}{\partial  {y}^\pm{}{}^\ga}{}\,\q\bp_\pm{}_\pa=
\frac{\partial}{\partial \bar{y}^\pm{}{}^\pa}\,.
\ee 
In these variables rank-two  covariant derivative  (\ref{Dtwsp8r})  is
\bee
\nn   D_2{}_{\mathfrak{u}}^{tw}&=&\dr   - \Big
(\go^{\ga\gb}y^+{}_\ga\p_+{}_\gb+
 \overline{\go}^{\pa\pb}\bar{y}^+{}_\pa \bp_+{}_\pb +
 \go^{\ga\gb}y^-{}_\ga \p_-{}_\gb+
 \overline{\go}^{\pa\pb}\bar{y}^-{}_\pa \bp_-{}_\pb  \Big )
 \\ \label{Dtwsp8r2}
&+&   f_{\ga\pa}(y^+{}^\ga{}\by^-{}^\pa+y^-{}^\ga{}\by^+{}^\pa)
+ h^{\ga\pa}(\p_+{}_\ga{}\bp_-{}_\pa+\p_-{}_\ga{}\bp_+{}_\pa)\\ \nn
&+&  \half b\big (4+ y^j{}{}^\ga \p_j{}_\ga
+\bar{y}^j{}{}^\pa \bp _j{}_\pa\big ) +  \half \widetilde{b}
\big ( y^ j{}{}^\ga \p_j{}_\ga
-\bar{y}^j{}{}^\pa \bp _j{}_\pa\big )
\q \eee
while
   operator  $D_2{}_{AdS}^{tw}$ (\ref{Dtwads}) results from $ D_2{}_{\mathfrak{u}}^{tw}$
   (\ref{Dtwsp8r2}) via  substitution
(\ref{adsc}).
Conserved currents $\PPP (y^\pm ,\by^\pm|x)$   satisfy  the
rank-two current equations \cite{gelcur}
 \be\label{ads2} D_2{}_{AdS}^{tw}\PPP(y^\pm ,\by^\pm|x)=0.\ee
           Evidently,  Eq.~(\ref{ads2}) decomposes into a set of subsystems
  \be\label{ads2n} D_2{}_{AdS}^{tw}\PPP_h(y^\pm ,\by^\pm|x)=0\q
  \Hh_2 \PPP_h(y^\pm ,\by^\pm|x)=h \PPP_h(y^\pm ,\by^\pm|x)\,\ee
   characterized by different eigenvalues of the rank-two helicity
 operator  $\Hh_2$  (\ref{hel1}).

 Deformed system (\ref{int}) in $AdS_4$ has the form \cite{GVC}
\bee\label{GeneralDefC}
D  C(y  ,\by |x)
+
\Big\{ e^{ \ga  }{}^{\pa}     y_\ga \F {}_{\pa}
 \PPP(y^\pm ,\by^\pm|x)\,
+
 e^{ \ga  }{}^{\pa}     \by_\pa
  \overline{\F}{}{}_{\ga}   {\III}(y^\pm ,\by^\pm|x)\Big\}
\,\Big|_{{y^\pm =\by^\pm =0}}  =0\,,
  \\ \nn \rule{0pt}{20pt}
D_2\PPP(y^\pm ,\by^\pm|x)\,=\,0\q  D_2\III(y^\pm ,\by^\pm|x)\, =\,0\,
 \qquad\qquad
\eee
   with   $D=D_1{}_{AdS}^{tw}$ (\ref{Dtwads}), $D_2= D_2{}_{AdS}^{tw}$   and
\bee\label{resultFpm1}
 \F_{\pa}&=&\sum_h \F_h{}_{\pa} \q\F_h{}_{\pa}=\left(F_h^+ \bp_+{}_{\pa}-F_h^- \bp_-{}_{\pa}\right)\q \\
 \label{resultFpm2}
  \overline{\F}_{\ga}&=&\sum_h \overline{\F}_h{}_{\ga}\,
\q \overline{\F}_h{}_{\ga}=\left(\overline{F _h{}^+} \p_+{}_{\ga}-\overline{F_h{}^- }\p_-{}_{\ga}\right)
\q  \eee\be
\label{resultFpm} F^\pm_{h}=\f{\p  }{\p \NNN_\pm } \sum_{n=0}^{2h}  {a}_{n, 2h}
 \big( \NNN_+\big)^{n }
   \big( \NNN_-\big)^{ { 2h-n}    }
  \sum_{k\ge0\, }
\f{\big(\overline{\NNN}_+\NNN_-+
\overline{\NNN}_-\NNN_+\big)^{k}}
 { k! (k+  2h+1 )!} ,\,\,\,\,\,   \NNN_\pm=y^\ga \p_\pm{}_\ga ,\,\,\,
\overline{\NNN}_\pm=\bar{y}^\pa \bp_\pm{}_\pa . \ee
Here  $\PPP$ and $\III$ can be independent rank-two fields.
Complex conjugated equations are analogous.

The coefficients $a_{n,2h}$ remain arbitrary.
Their absolute values reflect 
 the freedom in normalization of currents  of
different spins while phases can be understood as resulting from
EM-like duality transformations for different spins. Different phases
correspond to different models.  The freedom in such a phase  was originally observed
in \cite{Ann} in the analysis of HS interactions at the same order as in this paper.
However, since the analysis of \cite{Ann} respected HS symmetries, the latter expressed
 phases of different spins in terms of a single phase parameter
$\eta=\exp{i\varphi}$ which survives
 in the full nonlinear HS theory \cite{more}.

Note that the gluing operators $\F$ and $\overline{\F }$
in (\ref{GeneralDefC}) are such that
only components $\PPP_h$ of $\PPP$  carrying  rank-two helicities
 $h\ge -1$ contribute. Analogously,   only  $\III_h$  with $h\le 1$
 contribute to
 (\ref{GeneralDefC}).
For instance,   for rank-one fields of definite helicity
$\Hh_1 C_{\,h\,}= h\,C_{\,h\,},$ Eq.~(\ref{GeneralDefC}) yields for $h\geq0$
\bee\nn
 D_{AdS}^{tw} C_{\,h\,}(y,\by|x)
 + e^{ \ga  }{}^{\pa} y_\ga   \F_ {\,h\,} {}_{\pa}
  \PPP_{\,h-1\,}\,\Big|_{{y^\pm =\by^\pm =0}}  =0 \q
    \\\label{newtwCs+}
 D_{AdS}^{tw} C_{\,-h\,}(y,\by|x)
 + e^{ \ga  }{}^{\pa} \by_\pa
   \overline{\F}_{\,h\,}{}_{\ga} \III_{\,1-h\,}
 \,\Big|_{y^\pm =\by^\pm =0}
   =0
   \q
    \eee
\be\nn
D_{AdS}^{tw} C_{\,0\,}(y,\by|x) +e^{ \ga  }{}^{\pa} y_\ga \F_{\,0\,} {}_{\pa}
\PPP_{-1\,}\,\Big|_{y^\pm =\by^\pm =0}
 +     e^{ \ga  }{}^{\pa}     \by_\pa  \overline{\F} _{\,0\,} {}_{\ga} {\III} _{\,1\,} \,\Big|_{y^\pm =\by^\pm =0}
 =0
 \,
\ee
with $\F_s{}_{\pa}$ (\ref{resultFpm1}) and $\overline{\F}_s{}_{\ga}$   (\ref{resultFpm2}).

 The following simple  properties  are used below along with their complex conjugates:
 \be\label{properN}
 \big [      {A}\big(\NNN ,  \overline{\NNN}\big)  ,  {y}^j{}^\mu \big]=
 {y}^\mu\frac{\partial }{ \partial \NNN_j} {A}\big(\NNN ,  \overline{\NNN}\big)
  \q
  \Big[\frac{\partial }{ \partial {y}^\mu} , {A}{} \big(\NNN ,  \overline{\NNN}\big)\Big]
 =\frac{\partial }{ \partial \NNN_j} {A}\big(\NNN ,  \overline{\NNN}\big)
\frac{\partial }{ \partial {y}^j{}^\mu}
 \q
\ee\be\nn
  {A}\big(\NNN ,  \overline{\NNN}\big)\,
  y^k{}_\ga  {B} (y^\pm)\Big|_{{y^\pm=\by^\pm=0}}
     =y_\ga\f{\p   }{\p \NNN_k}
  \, {A}\big(\NNN ,  \overline{\NNN}\big)\,  {B}(y^\pm)
    \Big|_{{y^\pm=\by^\pm=0}}
    \quad \forall\,\,  {A},\,\,\,  {B}\,.\qquad \ee
As shown in \cite{GVC}, $  {F}_h^\pm$ (\ref{resultFpm}) obey
 \bee\label{Acontw++}
   { \f{\p  }{\p \NNN_-  } {F}_h^+\! \,+  \f{\p  }{\p \NNN_+  }{F}_h^-\! }=0
   \q
     {
\Big(   \f{\p^2  }{\p \NNN_+ \p \overline{\NNN}_-}+
 \f{\p^2 }{\p \NNN_- \p \overline{\NNN}_+}-1\Big){F}_h^\pm\!
  }&=&0
,\quad
  \\ \nn
 \Big\{2 + \NNN_k \frac{\partial }{\partial \NNN_k }
 \Big\}\Big( \frac{\partial {F}_h^+\!}{ \partial  \overline{\NNN}_-}\,
 +
  \frac{\partial {F}_h^-\!}{ \partial  \overline{\NNN}_+}\,
 \Big)
  - \NNN_- {F}_h^-\!- \NNN_+ {F}_h^+\!
  &=&0\,, \qquad\,
\\ \nn{ \Big\{2 + \NNN_k \frac{\partial }{\partial \NNN_k }
  \Big\}  \frac{\partial {F}_h^\pm\!}{ \partial  \overline{\NNN}_\pm}\,
   - \NNN_\mp {F}_h^\pm\!}
 &=&0 \,.\eee
$\overline F{}_h^\pm= \overline{{F}_h^\pm}$ obey  the conjugated relations.

\subsection{Conformal invariance of the deformation}
 To show that the $\mathfrak{sp}(4)$ invariant mixed system of \cite{GVC} is
 conformal consider deformed equation   (\ref{GeneralDefC})   with
  $D=D_{1}{}_{\mathfrak{su}}^{tw}$  ,
  $D_2=D_{2}{}_{\mathfrak{su}}^{tw}$ (\ref{Dtwcr}),
     $\overline{\F}_{\ga}$ (\ref{resultFpm1}) and  $ {\F}_{\pa}$
  (\ref{resultFpm2})\,.
For simplicity we set $\III=0$.
  Consistency  of these equations    restricts the deformation  by the conditions
      \bee \nn&&
 b h^{\gga\pga}
 y_\gga \F{}_{\pga}\PPP    (y^\pm \,,\bar{y}^\pm|x )\Big|_{{y^\pm=\by^\pm=0}}
+ b h^{\gga\pga}\Big\{\Big( 1 + \half\Big( y {}^\ga \p_\ga
+ \bar{y} {}^\pa \bp_\pa \Big)\Big)
 y_\gga \F{}_{\pga}
    \qquad\qquad\qquad\\ \nn&-&
  y_\gga \F{}_{\pga}\Big( 2+ \half\Big(y^+{}^\ga \p_+{}_\ga
+\bar{y}^+{}^\pa\bp_+{}_\pa+y^-{}_\ga \p_-{}_\ga
+\bar{y}^-{}^\pa \bp_-{}_\pa \Big)\Big)
    \Big\}\PPP    (y^\pm \,,\bar{y}^\pm|x )\Big|_{{y^\pm=\by^\pm=0}}
 \qquad\\ \label{Aadditconfconf}&+&
   h^{\gga\pga} h^{\ga\pb}
\Big\{\!  \p_\gga{}\bar{\p}_\pga
  y_\ga \F{}_{\pb}
 -  y_\ga \F{}_{\pb}
\Big(\p_-{}_\gga\bar{\p}_+{}_\pga+\p_+{}_\gga\bar{\p}_-{}_\pga\Big)
    \Big\}\PPP    (y^\pm \,,\bar{y}^\pm|x )  \Big|_{{y^\pm=\by^\pm=0}} \qquad
\\ &+&\nn  f^{\gga\pga} h^{\ga\pb}
 \Big\{y_\gga \bar{y}_\pga  y_\ga \F{}_{\pb}
- y_\ga \F{}_{\pb}   \Big(\by^+{}_\pga\,{y}^-{}_\gga+
\by^-{}_\pga\,{y}^+{}_\gga\,
\Big)  \,
    \Big\}\PPP\,   (y^\pm \,,\bar{y}^\pm|x )  \Big|_{{y^\pm=\by^\pm=0}}
   =0  \,.\qquad \eee
 Note that the first term in (\ref{Aadditconfconf})  results from $\mathrm{d}h^{\gga\pga}$ via
 flatness conditions (\ref{h4}), accounting for the conformal dimension of the frame
 field $h^{\gga\pga}$.

Using (\ref{resultFpm}), (\ref{properN}) along with  the decomposition
\be
\label{H} h^{\ga}{}^\pa \wedge h^\gb{}^\pb=\half\gvep^{\ga\gb}\overline {H}^{\pa\pb}
+\half\gvep^{\pa\pb}  {H}^{\ga\gb}\q
H^{\ga\gb} = h^{\ga}{}_\pa \wedge h^\gb{}^\pa\q
\overline {H}^{\pa\pb} = h_{\ga}{}^\pa\wedge h^{\ga\pb}\,,
\ee
the $h^2$ term in (\ref{Aadditconfconf}) demands
\bee\label{Acontw1.1}  H^{\pga \pb}
 \Big(\bar{\p}{}_\pga\,
 \Big\{2 + \NNN_k \frac{\partial }{\partial \NNN_k }
 \Big\}
- \NNN_+  \bar{\p}_-{}_\pga
 - \NNN_-  \bar{\p}_+{}_\pga
    \Big)\F{}_\pb
 &=&0\,, \qquad\, \\\nn H^{\mu\ga}
y_\ga   \Big(
\bar{\p}_\pga\,
\p_\mu
  -       (\p_-{}_\mu\bar{\p}_+{}_\pga+\p_+{}_\mu\bar{\p}_-{}_\pga)
    \Big)\F{}^\pga\,
 &=&0 \,.\qquad\eee
These conditions  hold true by virtue of (\ref{Acontw++}) along with (\ref{resultFpm1}).
The  $\,\vf h\, $ term  is also zero  by virtue of (\ref{Acontw++}) and (\ref{resultFpm1})
\bee\label{Aadditconfconff}2 \vf^{\gga\pga} h^{\ga\pb}
 y_\gga \bar{y}_\pga  y_\ga \Big\{1
- \Big( \f{\p}{\p \overline{\NNN}_-{} } \f{\p}{\p \NNN_+{} } +
\f{\p}{\p \overline{\NNN}_+{} } \f{\p}{\p \NNN_-{} }\Big) \Big\}
\F{}_{\pb}
    \PPP\,   (y^\pm \,,\bar{y}^\pm|x )  \Big|_{{y^\pm=\by^\pm=0}} =0\,.
\eee
   The $\,b h\,$ term   vanishes by virtue of (\ref{properN})
\be\label{bh}  {b} h^{ \gga\pga}y_\gga\Big\{
1+
\half \Big(  3 +  \NNN_k
\frac{\partial}{\partial \NNN_k}
+\overline{\NNN}_k\frac{\partial}{\partial \overline{\NNN}_k}\Big) - \half \Big(5+
\NNN_k   \frac{\partial}{\partial \NNN_k}
+\overline{\NNN}_k\frac{\partial}{\partial \overline{\NNN}_k}\Big)\Big\}
\F{}_{\pga}
   \PPP\,\Big|_{{y^\pm=\by^\pm=0}}
  =0.\qquad \ee
Here it is important that the first term accounting  for  the conformal
dimension of  $h^{\gga\pga}$, precisely compensates the difference between
the rank-one and rank-two vacuum  conformal dimensions.

 This proves consistency of the conformal deformation of  Eq.~(\ref{GeneralDefC}) with $D=D_1{}_{\mathfrak{su}}^{tw}$  and
 $D_2=D_2{}_{\mathfrak{su}}^{tw}$    (\ref{Dtwcr}) hence
 implying  conformal invariance of this system. 

\subsection{Inconsistency of the  $\mathfrak{u}(2,2)$ extension  }
\label{inconsistency}
Now we are in a position to show that the current deformation is
not $\mathfrak{u}(2,2) $ invariant.

Consider Eq.~(\ref{GeneralDefC})   with 
$D=D_1{}_{\mathfrak{u}}^{tw}$,
 $D_2=D_2{}_{\mathfrak{u}}^{tw}$      (\ref{Dtwsp8r}) and  $F^j$ (\ref{resultFpm}).
In addition to Eq.~(\ref{Aadditconfconf}),
 its  consistency demands  by  virtue of (\ref{properN}) that
 \be\label{btilde-e} \,\,  \widetilde{b} h^{ \ga\pb}y_\ga\Big\{\half
 \Big(  1+  \NNN_k   \frac{\partial}{\partial \NNN_k}
-\overline{\NNN}_k\frac{\partial}{\partial \overline{\NNN}_k}\Big)
  - \half\Big(-1+  \NNN_k   \frac{\partial}{\partial \NNN_k}
-\overline{\NNN}_k\frac{\partial}{\partial \overline{\NNN}_k}\Big)
  \Big\}\F{}_{\pb}
 \PPP\Big|_{{y^\pm=\by^\pm=0}}\quad\ee\be \nn+
 \widetilde{b} h^{ \ga\pb}\by_\pb\Big\{\half\Big(  -1+  \NNN_k   \frac{\partial}{\partial \NNN_k}
-\overline{\NNN}_k\frac{\partial}{\partial \overline{\NNN}_k}\Big)
  - \half\Big( 1+  \NNN_k   \frac{\partial}{\partial \NNN_k}
-\overline{\NNN}_k\frac{\partial}{\partial \overline{\NNN}_k}\Big)
  \Big\}\overline{\F}_{\ga}
  {\III}\,\Big|_{{y^\pm=\by^\pm=0}}\qquad\,\,
    \ee
should be zero. Here all terms cancel except for the vacuum contributions
which do not  because the  helicity operator  counts the difference between
powers  of   $y$ and $\by$ while $y_\ga F^\pm \bp_\pm{}_{\pb}$ is proportional to
$y\f{\p}{\p \by^\pm}$.  (Note that helicity of the frame field is zero.)
As  a result,   (\ref{btilde-e}) takes the form
\be\label{F-F}\widetilde{b} h^{ \ga\pb}\Big(y_\ga   \F_{\pb}
   \PPP\, - \by_\pb   {\overline{\F}} _{\ga}
       {\III}\Big)\Big|_{{y^\pm=\by^\pm=0}} \,.\ee
This is  however nonzero.
Indeed, decomposition (\ref{newtwCs+})
brings $\tilde b h$ term (\ref{F-F}) to the form
\be\label{F-Fd}\widetilde{b} h^{ \ga\pb}\Big\{\sum_{h>0}\Big(y_\ga   \F_h {}_{\pb}
   \PPP_{h-1}\, - \by_\pb   \overline{\F} _h  {}_{\ga}
    {\III}_{1-h}\Big)+
    y_\ga \F_{\,0\,} {}_{\pb} \PPP_{-1\,}\,
 -         \by_\pb  \overline{\F} _{\,0\,}{}_{\ga} {\III} _{\,1\,}
        \Big\}\Big|_{{y^\pm=\by^\pm=0}} \,.\ee
Since the terms $\PPP_{h-1}$ and $\III_{1-h}$ are independent for $h>0$,
they should vanish separately which would imply that the current deformation
is trivial. For $h=0$ the two terms have different total (\ie rank-one)
helicity and hence, again, should vanish separately.

An attempt to  compensate the   $\widetilde{b} h $-term  in (\ref{btilde-e})
 by  an additional $\widetilde{b}$-dependent deformation term $\widetilde{b} G
\widetilde{\PPP}\,\Big|_{{y^\pm=\by^\pm=0}} $ for some operator
  $G $ and rank-two  field $\widetilde{\PPP}$ fails because the modified consistency
  condition then would imply
\bee\label{btildedop}
\widetilde{b} h^{ \ga\pb}\Big(y_\ga   \F {}_{\pb}
   \PPP\, - \by_\pb   {\overline{\F}}{}_{\ga}
   {\III}\Big)\Big|_{{y^\pm=\by^\pm=0}}
=   \widetilde{b} D{}{}_{\mathfrak{u}}^{tw} G \widetilde{\PPP}\,\Big|_{{y^\pm=\by^\pm=0}} \equiv
\widetilde{b} D{}{}_{\mathfrak{su}}^{tw} G
\widetilde{\PPP}\,\Big|_{{y^\pm=\by^\pm=0}} \,. \eee
However if (\ref{btildedop})  admitted a solution, by
$\tilde  b \wedge\tilde b=0$ (\ref{btb1})
this would imply that
 the original $\mathfrak{su}(2,2) $ current deformation $h^{ \ga\pb}y_\ga   \F {}_{\pb}
  \PPP\,\Big|_{{y^\pm=\by^\pm=0}}+ h^{ \ga\pb}\by_\pb   \overline{\F} {}_{\ga}
  \III\,\Big|_{{y^\pm=\by^\pm=0}}$  is trivial which is not
  the case.

Thus it is shown that the current deformation of free field equations
does not allow a  $\mathfrak{u}(2,2) $-symmetric extension of the
 conformal deformation. Since the helicity operator $\Hh$
(\ref{hel1}) is a Cartan element of $\mathfrak{sp}(8)$, this implies that
the $\mathfrak{sp}(8)$ invariant form of the current interactions
is  ruled out as well, \ie HS current interactions
necessarily break  $\mathfrak{sp}(8)$ down to $\mathfrak{su}(2,2)$.

\section{Discussion}

The current deformation of the HS equations is shown to break
the $\mathfrak{sp}(8)$ symmetry of the free massless field equations
 down to its conformal subalgebra. The analysis is performed
in terms of the gauge invariant field strengths.
To rule out the full  $\mathfrak{sp}(8)$ symmetry  it was enough to show
the $\mathfrak{u}(1)$
helicity symmetry, which is a part of  $\mathfrak{sp}(8)$, is inconsistent with the
current interactions. This is not too surprising since the $\mathfrak{u}(1)$ helicity
symmetry describes EM duality transformations
known to be broken by the current interactions.

Our conclusions are in agreement with the structure of nonlinear HS field equations of
\cite{more} which contain a free phase parameter
$\eta=\exp i\varphi$  forming a one-dimensional representation of
the $\mathfrak{u}(1)$ helicity transformations that relates inequivalent HS
theories associated with different values of $\varphi$. Hence
the symmetry $\mathfrak{u}(1)$ of EM duality
 is not a symmetry of a  HS theory
with given $\eta$, mapping one theory to another.
As such the parameter $\eta$ is reminiscent of
the embedding tensor introduced in \cite{deWit:2002vt,de Wit:2007mt} to describe different models
of supergravity. It would be also interesting to check its possible relation with
the parameter of $\go$-deformation in supergravity
in the context of ABJM   theory \cite{Borghese:2014gfa}
argued to play an important role in HS holography \cite{Chang:2012kt}.

A useful  viewpoint is to treat $\eta$ as a  VEV 
for some field affected by the  symmetry $\mathfrak{u}(1)$ of EM duality.
It would be interesting to look for a Higgs-like field
in the  nonlinear HS theory of \cite{more} that would break $\mathfrak{sp}(8)$
down to the conformal algebra allowing to treat the $\mathfrak{sp}(8)$ symmetry
of the HS theory as spontaneously broken. Further extension of these ideas
can lead to better insight into possible origin of
duality symmetries in HS theory in spirit of the discussion of
\cite{Sezgin:1998gg}

Note that being described in terms
of a doubled set of oscillators $z_\ga, y_\ga, \bar z_{\dot \ga}, \bar y_{\dot \ga}$,
the nonlinear equations of \cite{more} do have a spontaneously broken
$\mathfrak{sp}(8)$ symmetry. A possible relation between the two
$\mathfrak{sp}(8)$ symmetries is not direct however because the one associated with the
additional oscillators $z_\ga, \bar z_{\dot \ga}$ does not act properly  on
the free fields while that broken by the current interactions does.

Also let us stress that the proof of conformal invariance of the current interactions
was given in the sector of  0-forms associated with the gauge
invariant HS curvatures. We do not expect  it to be extendable to the sector of
gauge fields in the standard setup. (Recall that linearized Einstein equations are
not conformal invariant in terms of the metric tensor while their consequences
 for the Weyl
tensor are.) This can probably be achieved, however, in $AdS_4$ using the approach of
\cite{33}.

It should be mentioned that the $\mathfrak{sp}(8)$-invariant formalism extended
at the level of free fields to HS 1-form connections in \cite{33} contains
a doubled set of all fields including the gauge fields.
In this approach EM duality  rotates the fields in the doublet. As is well known,
 for a doubled set of fields EM duality is much easier to achieve
 \cite{Zwan,Deser:1976iy,SS,pst,PST1}. Hence one might expect that the
 obstruction of this paper can be avoided in the setup of \cite{33}.
Unfortunately we were not able to proceed along these lines.
Basically the difficulty remains the same as in this paper if the current
is constructed from the 0-form rank fields of the same type, \ie
$J_{ii}\sim C_{ii}C_{ii} $ with $i=0,1$ in notations of \cite{33},
while other options  do not have much sense at all from the perspective
of matching of the left and right hand sides of the deformed equations.

Another interesting point is that,
as shown in \cite{Deser:1976iy}
(see also \cite{Bunster:2010wv} and references therein),
within the Hamiltonian-like approach breaking manifest Lorentz symmetry,
EM
duality can be realized as a manifest symmetry even at the Lagrangian level.
The negative conclusion on the possibility to preserve
manifest duality invariance was achieved in this paper in the manifestly Lorentz covariant
approach, \ie it was shown that the maximal subalgebra of $\mathfrak{sp}(8)$
that contains the Lorentz subalgebra $\mathfrak{sl}(2|\mathbb{C})$
and admits a consistent current deformation is
$\mathfrak{su}(2,2)$ which does not contain the EM duality
generator. The remaining question  is whether there exists a subalgebra
$\mathfrak{h}\in \mathfrak{sp}(8)$ that does not contain  $\mathfrak{sl}(2|\mathbb{C})$
but contains the helicity generator $\Hh$ and respects some current-like deformation.

 \section*{Acknowledgment}

We are grateful to Nicolas Boulanger, Simone Giombi, Stamatios Nicolis,
Ergin Sezgin, Evgeny Skvortsov, Per Sundell and  Dmitry Sorokin
for useful comments.
This research was supported in part by the RFBR Grant No 14-02-01172.

\newcounter{appendix}
\setcounter{appendix}{1}
\renewcommand{\theequation}{\Alph{appendix}.\arabic{equation}}
\addtocounter{section}{1} \setcounter{equation}{0}
 \renewcommand{\thesection}{\Alph{appendix}.}
 \addtocounter{section}{1}
\addcontentsline{toc}{section}{\,\,\,\,\,\,\,Appendix.  Inequalities}

\section*{Appendix.  Inequalities}
\newcommand{\stt}{{\star}}
 \newcommand{\bs}{{\bar s}}
\newcommand{\gt}{{\tau}}
 The action of $D$ on the $\go$-dependent terms in (\ref{int1})
gives by virtue of (\ref{dg})
\be
\label{1}
G(w,C)\star C - C\star \widetilde G(w,C)\,.
\ee
In the linearized approximation $G(w,C)$ is
\be
G(w,C)(y,\bar y) = \eta \overline{H}^{\pga\pb} \f{\p^2}{\p \bar y^{\pga} \p \bar y^\pb}
C(0,\bar y) + \bar \eta {H}^{\ga\gb} \f{\p^2}{\p  y^{\ga} \p  y^\gb}
C(y, 0)\,,
\ee
where $\eta$ is a complex parameter.
For given spins $s_1$ and $s_2$ of the first and second factors of $C$
the $\eta$-dependent part of the first term of (\ref{1}) is
proportional to
\be\label{Ctot}
  X = \int d \bar s\,d \bar t
 \,\exp\big(i\,\bar s_\pa\,\bar t^\pa\big) \overline{H}^{\pa\pb}\bt_\pa \bt_\pb
 C_{s_1}(0,\by+\bs) C_{s_2}(y, \by+ \bt)\,.
 \ee
This gives
\be\label{spins s1s2}
2s_1=\overline{N}_{\by_1}+\overline{N} \q
2s_2=\big|-{N}_{ y }+\overline{N}_{\by_2}+\overline{N} -2 \big|\q
 2s = \big|-{N}_{ y }+\overline{N}_{\by_2}+\overline{N}_{\by_1 } \big|\,,
\ee
where $s$ is a spin of $X$,  $N_y$ is the degree of $y$, $\overline{N}_{\bar y_1}$ and $\overline{N}_{\bar y_2}$ are the degrees
of $\bar y$ in the first and second factors of $C$, respectively, while
$ \overline{{N}}$ is the degree of the integration parameter of $\bar s$ equal to that of
$\bar t$.
Since $ \overline{H}^{\pa\pb}\bt_\pa \bt_\pb$ has  degree two in $\bar{t}$
\be\label{Ngt2}\overline{N}\ge2\,.\ee

The main fact is  that $ X$ can be nonzero only if
 \be\label{ineqss1s2}s  < s_1+s_2\,.
 \ee

The proof is straightforward.

 For
\be\label{11}
{N}_{ y }\ge\overline{N}_{\by_2}+\overline{N} -2\,\q{N}_{ y }\le\overline{N}_{\by_2}
+\overline{N}_{\by_1}
\ee
(\ref{spins s1s2}) yields
\be\label{s11}2s_1=\overline{N}_{\by_1}+\overline{N}  \q  2s_2=
{N}_{ y }-\overline{N}_{\by_2}-\overline{N} +2\q
2s=\overline{N}_{\by_1}+\overline{N}_{\by_2}-N_y.\ee
Assuming that $s \ge s_1+s_2$
one has from (\ref{s11})
\be\label{ineq11}
\overline{N}_{\by_1}+\overline{N}_{\by_2}-N_y>\overline{N}_{\by_1}+\overline{N} +  {N}_{ y }-\overline{N}_{\by_2}-\overline{N} \q
 \overline{N}_{\by_2}\ge N_y+1\q\ee
that by virtue of (\ref{11}) yields $\overline{N}\le1$, in  contradiction with
 (\ref{Ngt2}).

 For
\be\label{12}
{N}_{ y }\ge\overline{N}_{\by_2}+\overline{N}-2 \,\q{N}_{ y }\ge\overline{N}_{\by_2}+\overline{N}_{\by_1}
\ee
Eq.~(\ref{spins s1s2}) yields
\be\label{s12}2s_1=\overline{N}_{\by_1}+\overline{N}  \q
2s_2=  {N}_{ y }-\overline{N}_{\by_2}-\overline{N} +2\q
2s=-\overline{N}_{\by_1}-\overline{N}_{\by_2}+N_y.\ee
Assuming that $s \ge s_1+s_2$ (\ref{s12}) leads to the contradiction
$
  -\overline{N}_{\by_1}\ge 1\,.
$

 For
\be\label{21}
{N}_{ y }\le\overline{N}_{\by_2}+\overline{N}-2 \,\q{N}_{ y }\le\overline{N}_{\by_2}+\overline{N}_{\by_1}
\ee
Eq.~(\ref{spins s1s2}) yields
\be\label{s21}2s_1=\overline{N}_{\by_1}+\overline{N} \q
2s_2= - {N}_{ y }+\overline{N}_{\by_2}+\overline{N} -2 \q
2s=\overline{N}_{\by_1}+\overline{N}_{\by_2}-N_y.\ee
For $s \ge s_1+s_2$ this yields $\overline{N}\le1$, in  contradiction with
 (\ref{Ngt2}).

For
\be\label{22}
{N}_{ y }\le\overline{N}_{\by_2}+\overline{N}-2 \,\q{N}_{ y }\ge\overline{N}_{\by_2}+\overline{N}_{\by_1}
\ee
(\ref{spins s1s2}) yields
\be\label{s22}2s_1=\overline{N}_{\by_1}+\overline{N}  \q
2s_2= - {N}_{ y }+\overline{N}_{\by_2}+\overline{N} -2 \q
2s=-\overline{N}_{\by_1}-\overline{N}_{\by_2}+N_y.\ee
At $s \ge s_1+s_2$ this gives
$
 N_y\ge \overline{N}_{\by_1}+\overline{N}  +\overline{N}_{\by_2}-1
$
in contradiction with  (\ref{22}). This finishes the proof of
inequality (\ref{ineqss1s2}). The proof for the other terms in (\ref{1}) is
analogous.

\end{document}